\documentclass[reqno,12pt]{amsart} %{article}
\usepackage{epsf}\usepackage{amssymb} \usepackage{cite}
\usepackage{amsthm}
\usepackage{amscd}
\usepackage{amsfonts}
\usepackage{amsmath}

\newcommand{\be}{\begin{equation}}
\newcommand{\ee}{\end{equation}}
\newcommand{\bea}{\begin{eqnarray}}
\newcommand{\eea}{\end{eqnarray}}

\usepackage{hyperref}
\def\stackreb#1#2{\ \mathrel{\mathop{#1}\limits_{#2}}}

\numberwithin{equation}{section}
\newcommand{\C}{\mathbb C}

\newcommand{\R}{\mathbb R}
\newcommand{\Z}{\mathbb Z}

\newcommand{\ve}{\varepsilon}

\begin{document}

\title[The elliptic hypergeometric function and $6j$-symbols for SL(2,$\C)$  group]
{The elliptic hypergeometric function \\[1mm] and $6j$-symbols for the SL(2,$\C$) group}

\author{S. E. Derkachov, G. A. Sarkissian and V. P. Spiridonov}

\makeatletter
\renewcommand{\@makefnmark}{}
\makeatother
\footnote{\small
This study has been partially funded within the framework of the HSE University
Basic Research Program and by the Russian Science Foundation (project no. 19-11-00131)
}

\address{
S.D., G.S., V.S.: St. Petersburg Department of the Steklov Mathematical Institute
of Russian Academy of Sciences, Fontanka 27, St. Petersburg, 191023 Russia
\linebreak
\indent
G.S., V.S.: Laboratory of Theoretical Physics, JINR, Dubna, 141980,  Russia
\linebreak
\indent
G.S.: Yerevan Physics Institute, Alikhanian Brothers 2, 0036 Yerevan, Armenia
\linebreak
\indent
 V.S.: Laboratory for Mirror Symmetry, NRU HSE, Moscow, Russia}

\begin{abstract}
We show that the complex hypergeometric function describing $6j$-symbols for $SL(2,\C )$ group
is a special degeneration of the $V$-function --- an elliptic analogue of the Euler-Gauss
$_2F_1$ hypergeometric function. For this function, we derive mixed difference-recurrence relations
as limiting forms of the elliptic hypergeometric equation and some symmetry transformations.
At the intermediate steps of computations, there emerge a function describing the $6j$-symbols for the
Faddeev modular double and the corresponding difference equations and symmetry transformations.
\end{abstract}

\maketitle
%{\bf Keywords: $6j$-symbols, $SL(2,\mathbb{C})$ group, elliptic hypergeometric function}

\tableofcontents

\section{Introduction}

Classical special functions \cite{aar} have natural interpretation as matrix elements
of the operators realizing representations of standard Lie groups.
For example, matrix elements of the $SL(2,\mathbb{R})$ group representations are described
by the Euler-Gauss $_2F_1$ hypergeometric function  \cite{vil}.
% In the representation theory of Lie groups, there is a problem
One of the problems in the representation theory of Lie groups is
% of
the decomposition of tensor products of irreducible
representations into a direct sum of irreducible representations. In the simplest
case of the product of two representations the expansion coefficients are called
Clebsch-Gordan coefficients, or $3j$-symbols, which are again related to classical special
functions. For triple tensor products, there are two different ways of sequential pairwise tensoring
of representations leading to two different expressions for the expansion coefficients in terms
of $3j$-symbols.
Matrix elements of the map between these two possible expansions are called $6j$-symbols \cite{VMK}
and play a very important role both in physics and mathematics. In particular,
they have found applications in quantum mechanics, two-dimensional
conformal field theory, three-dimensional gravity, solutions of the Yang-Baxter equation
and related integrable systems, knot theory, topology, and so on.

The well-known Racah polynomials were introduced as $6j$-symbols of the $SU(2)$ group \cite{Racah}.
Their explicit expression uses a special (Saalsch\"utzian) terminating hypergeometric
$_4F_3$-series. Similar situation
holds for the $q$-Racah and Askey-Wilson polynomials \cite{aw}: they are related to $6j$-symbols
of the quantum algebra $sl_q(2,\mathbb{R})$ \cite{KR,Groenevelt} and are explicitly described
by terminating $q$-hypergeometric $_4\varphi_3$-series.

The most complicated form of $6j$-symbols emerges for non-compact groups and their
principal-series representations. Our key object here is the $6j$-symbols for the
group $SL(2,\mathbb{C})$ (or proper Lorentz group $SO(3,1)^+$ isomorphic to it),
constructed in \cite{Ismag2} and \cite{Derkachev:2019bcl}.
They are expressed in terms of a three complex-dimensional integral (six real
dimensions), or, equivalently, as an infinite bilateral sum of the Mellin-Barnes
type integrals. Another group-theoretical object that we are interested in the present paper is
the Faddeev modular double $sl_q(2,\mathbb{R})\times sl_{\tilde q}(2,\mathbb{R})$
\cite{fad:mod}. Its continuous-series $6j$-symbols are
built in \cite{Ponsot:2000mt} and have the form of a contour
integral of a combination of Faddeev's modular quantum dilogarithms \cite{Fad94,Fad95}.
This function is directly
related to the hyperbolic hypergeometric function introduced by Ruijsenaars in \cite{RuijAW},
which can be considered a hyperbolic analogue of the Askey-Wilson function
since both satisfy one finite-difference equation.

The next level of generalization of the $6j$-symbols is associated with the elliptic
hypergeometric functions. Such objects appeared first in the context of
elliptic solutions of the IRF (interaction round a face) type Yang-Baxter
equation, which combine into an elliptic function
taking the form of a terminating elliptic hypergeometric series \cite{ft}. The general $V$-function --- a genuine
elliptic hypergeometric function that is transcendental over the field of elliptic
functions and which absorbs the elliptic $6j$-symbols of \cite{ft} --- was constructed in \cite{spi:theta}.
This $V$-function represents an elliptic analogue of the Euler-Gauss hypergeometric function
satisfying a second order difference equation with elliptic coefficients,
which is called the elliptic hypergeometric equation \cite{spi:thesis,spi:tmf}.

As shown in \cite{Sarkissian:2020ipg}, the elliptic beta integral \cite{spi:umn} and $V$-function
can be reduced to rather general complex hypergeometric functions in the Mellin-Barnes
representation (with the hyperbolic hypergeometric functions obtained at the intermediate steps).
In this paper, we show that the Mellin-Barnes form of $6j$-symbols obtained in
\cite{Ismag2,Derkachev:2019bcl} is a special subcase of a complex hypergeometric function
constructed independently in \cite{DM} and \cite{Sarkissian:2020ipg}.
We give also a detailed comparison of these $6j$-symbols with the functions emerging from $b\to \textup{i}$
limit of the $6j$-symbols for the Faddeev modular double \cite{Ponsot:2000mt}.

In addition to the $b\to \textup{i}$ limit for the hyperbolic integral of Ruijsenaars \cite{RuijAW}
associated with the $6j$-symbols of \cite{Ponsot:2000mt},
we describeits $b\to 0$ limit in some detail and derive corresponding identities.
In particular, we derive difference equations satisfied by the limiting
hypergeometric functions in different normalizations.
In \cite{GSVS0}, a new singular limit $b\to 1$ was found for the Faddeev modular dilogarithm.
It can be applied to the same Ruijsenaars integral \cite{RuijAW} in order to derive
symmetry relations for a particular rational hypergeometric function emerging in this limit.
However, we skip consideration of the corresponding identities in this paper.
our general conclusion here is that the complex $6j$-symbols, together with many other
generalized hypergeometric functions, are degenerations of the elliptic
hypergeometric $V$-function, which confirms its status of a universal special function
of hypergeometric type.

\section{Complex $6j$-symbols}

The well known Euler's beta integral has the form
$$
\int_0^1x^{\alpha-1}(1-x)^{\beta-1}dx =\frac{\Gamma(\alpha)\Gamma(\beta)}{\Gamma(\alpha+\beta)},
\quad \text{Re}(\alpha),  \text{Re}(\beta)>0,
$$
where  $\Gamma(x)$ is the Euler gamma function.
Its generalization to the field of complex numbers $x\to z\in\C$ was suggested in \cite{GGV}.
Its description requires the complex gamma function
\begin{equation}
{\bf \Gamma}(x,n)={\bf \Gamma}(\alpha|\alpha'):=\frac{\Gamma(\alpha)}{\Gamma(1-\alpha')}
=\frac{\Gamma(\frac{n+\textup{i}x}{2})}{\Gamma(1+\frac{n-\textup{i}x}{2})},
\quad \alpha=\frac{n+\textup{i}x}{2},\; \alpha'=\frac{-n+\textup{i} x}{2},
\label{Cgamma}\end{equation}
where $x\in \C $ and $n\in\Z$. For two complex numbers  $\alpha, \alpha'\in\mathbb{C}$
such that $\alpha-\alpha'=n  \in\mathbb{Z}$, we use the notation
\begin{equation}
[z]^\alpha:= z^\alpha \bar z^{\alpha'}=|z|^{2\alpha'} z^{n},\quad
\int_{\mathbb{C}} d^2z:=\int_{\mathbb{R}^2}d(\text{Re}\, z)\, d(\text{Im}\, z),
\label{[z]}\end{equation}
where $\bar z$ is the complex conjugate of $z$. Then the complex beta integral in \cite{GGV}
can be represented, after a linear fractional transformation of the integration variable,
in the form of the star-triangle relation
\begin{equation}
\int_{\mathbb{C}}[z_1-w]^{\alpha-1} [z_2-w]^{\beta-1} [z_3-w]^{\gamma-1}\frac{d^2w}{\pi}
=\frac{ {\bf\Gamma}(\alpha,\beta,\gamma) } {[z_3-z_2]^{\alpha}[z_1-z_3]^{\beta}[z_2-z_1]^{\gamma}},
\label{str}\end{equation}
where ${\bf\Gamma}(\alpha_1,\ldots,\alpha_k)
:=\prod_{j=1}^k{\bf\Gamma}(\alpha_j|\alpha_j')$ and $\alpha+\beta+\gamma= \alpha'+\beta'+\gamma'=1$.

The right-hand side expression in \eqref{str} can take different forms due to the
reflection equations
\be
{\bf \Gamma}(\alpha|\alpha') =(-1)^{\alpha-\alpha'}{\bf \Gamma}(\alpha'|\alpha), \qquad
{\bf \Gamma}(x,-n)=(-1)^n{\bf \Gamma}(x,n),
\label{reflCgamma0}\ee
and
\be
{\bf \Gamma}(\alpha|\alpha'){\bf \Gamma}(1-\alpha|1-\alpha')  =(-1)^{\alpha-\alpha'}, \qquad
{\bf \Gamma}(x,n){\bf \Gamma}(-x-2\textup{i},n)=1.
\label{reflCgamma}\ee
The Mellin-Barnes form of identity \eqref{str}  was derived in \cite{DMV2017}.
A systematic consideration of complex hypergeometric functions of one variable is given
in \cite{Neretin}, and we partially employ the notations introduced in that paper.

The $3j$-symbols for  unitary principal series representations of the $SL(2,\C)$ group were computed
by Naimark long ago  \cite{Naimark}. $6j$-symbols for such representations
$\mathrm{R}_{\ell}(c,{\tilde c} )$ were first calculated
by Ismagilov in \cite{Ismag2} and were later they rederived in \cite{Derkachev:2019bcl},
where the method of computations of Feynman diagrams was used for solving an integral equation
for these $6j$-symbols. Initially they were represented as a triple
complex integral (i.e., a 6-dimensional integral over real variables) of elementary functions,
\be
\mathrm{R}_{\ell}(c,{\tilde c} ) = \int_\C \mathrm{d}^2z\,
\Psi_2(a_1,a_2,a_3|\ell,c,z)
\overline{\Psi_1(a_1,a_2,a_3|\ell,{\tilde c},z)},
\label{complex6j}\ee
where
$$
\overline{\Psi_1(a_1,a_2,a_3|\ell,{\tilde c},z)} =
\int_\C \frac{\mathrm{d}^2y}
{[1-y]^{\frac{1-\ell-{\tilde c} -a_3}{2}}
[y]^{\frac{1+\ell-{\tilde c} +a_3}{2}}
[z-y]^{\frac{1-a_1+a_2+{\tilde c} }{2}}}
$$
and
$$
\Psi_2(a_1,a_2,a_3|\ell,c,z) =\frac{1}{[z]^{\frac{1+a_1-\ell+c}{2}}}
\int_\C \frac{\mathrm{d}^2z_0}
{[1-z_0]^{\frac{1+a_1+\ell+c}{2}}
[z_0]^{\frac{1+a_2-a_3-c}{2}}
[z-z_0]^{\frac{1-a_2+a_3-c}{2}}}.
$$
The principal series representation parameters $a_1, a_2, a_3, l, c ,\tilde c $ are of the form $\alpha=(n_\alpha+\textup{i}x_\alpha)/2,$ where $n_\alpha\in\Z,\, x_\alpha\in\R$
are such that linear combinations of integers $n_\alpha$ appearing in the powers
of square bracket expressions \eqref{[z]} take integer values only, so that all resulting functions are
single-valued.

This 6-dimensional integral can be rewritten as an infinite bilateral sum of integrals corresponding to the Mellin-Barnes
type representation of the function of interest. Such an expression was obtained in \cite{Ismag2} and slightly corrected
in \cite{Derkachev:2019bcl}:
\bea\nonumber &&
R_l(c,\tilde c )=(-1)^{\tilde c -\tilde c'}{\pi^2\over 4}
{a\left({1-a_3-l+\tilde c \over 2}\right)a\left({1+a_1+l+c\over 2}\right)\over a\left({1+a_1-a_2+\tilde c \over 2}\right)a\left({1+a_2-a_3+c\over 2}\right)}
\\ && \makebox[-2em]{} \times
\sum_{n\in\Z}\int_L {a\left({1+a_1-a_2+\tilde c \over 2}+s\right)a\left({1-a_1-a_2+\tilde c \over 2}+s\right)a\left({1+a_3+l+\tilde c \over 2}+s\right)a\left({1-a_3+l+\tilde c \over 2}+s\right)\over
a(s)a(\tilde c +s)a\left({-c-a_2+l+\tilde c \over 2}+s\right)a\left({c-a_2+l+\tilde c \over 2}+s\right)}\, du,
\label{6jds} \eea
where $s=(n+\textup{i} u)/2$ and
\be\label{alphon}
a(\alpha)={\Gamma(1-\alpha')\over \Gamma(\alpha)}.
\ee
The integration contour $L$ may be any contour lying
in the strip Im$(u)\in]-1,0[$ (singularities of the integrand lie on the real axis due to the unitarity condition $x_\alpha\in\R$).
Expression \eqref{6jds} differs from the one in \cite{Ismag2} by the sign of the parameter $\tilde c$,
corresponding to the transition to an equivalent representation, which we believe is the correct prescription.

Now we rewrite the complex $6j$-symbols in terms of the complex gamma function (\ref{Cgamma}).
If we set  $\alpha=N/2+\textup{i}\sigma/2$ and $\alpha'=-N/2+\textup{i}\sigma/2$ in (\ref{alphon}), we obtain
\be\label{algam}
a(\alpha)={1 \over {\bf \Gamma}(\sigma,N)}={\bf \Gamma}(-\sigma-2\textup{i},N).
\ee
Using relation (\ref{algam}) and setting
\begin{equation}
\begin{aligned}
&a_1=N_1/2+\textup{i}\sigma_1, \\
&a_2=N_2/2+\textup{i}\sigma_2,
\end{aligned}\qquad
\begin{aligned}
&a_3=N_3/2+\textup{i}\sigma_3,\\
&l=N_4/2+\textup{i}\sigma_4,
\end{aligned}\qquad
\begin{aligned}
&c=M_1/2+\textup{i}\rho_1,\\
&\tilde c =M_2/2+\textup{i}\rho_2,
\end{aligned}\qquad
\begin{aligned}
&s=N/2+\textup{i}u/2,
\end{aligned}
\end{equation}
we rewrite expression (\ref{6jds}) as
\bea \nonumber && \makebox[-1em]{}
\big\{\,{}^{\sigma_1,N_1}_{\sigma_3,N_3}\,{}^{\sigma_2,N_2}_{{\sigma}_4,N_4}\,|\,{}^{\rho_1,M_1}_{\rho_2,M_2}\big\}
={\pi^2\over 4}{{\bf \Gamma}\left(\sigma_1-\sigma_2+\rho_2-\textup{i}, A_1\right)
{\bf \Gamma}\left(\sigma_2-\sigma_3+\rho_1-\textup{i}, A_2\right)\over {\bf \Gamma}\left(-\sigma_3-\sigma_4+\rho_2-\textup{i}, A_3\right)
{\bf \Gamma}\left(\sigma_1+\sigma_4+\rho_1-\textup{i}, A_4\right)}
\\ && \makebox[-2em]{} \times
(-1)^{M_2-N_2+N_4}\sum_{N\in\Z}\int_{u\in L}
\prod_{j=1}^4 {\bf \Gamma}(R_j-u,S_j-N)
{\bf \Gamma}(U_j+u,T_j+N)\, du,
\label{idsgamma}\eea
where
\begin{equation}\label{RU}
\begin{aligned}
&R_1=-\sigma_1+\sigma_2-\rho_2-\textup{i},\\
&R_2=\sigma_1+\sigma_2-\rho_2-\textup{i},\\
&R_3=-\sigma_3-\sigma_4-\rho_2-\textup{i},\\
&R_4=\sigma_3-\sigma_4-\rho_2-\textup{i},
\end{aligned}\quad
\begin{aligned}
&U_1=-\rho_1-\sigma_2+\sigma_4+\rho_2,\\
&U_2=\rho_1-\sigma_2+\sigma_4+\rho_2,\\
&U_3=0,\\
&U_4=2\rho_2,
\end{aligned}\quad
\begin{aligned}
&S_1=(-N_1+N_2-M_2)/2,\\
&S_2=(N_1+N_2-M_2)/2,\\
&S_3=-(N_3+N_4+M_2)/2,\\
&S_4=(N_3-N_4-M_2)/2
\end{aligned}
\end{equation}
and
\begin{equation}\label{T}
T_1=(-M_1-N_2+N_4+M_2)/2, \quad T_2=(M_1-N_2+N_4+M_2)/2, \quad T_3=0, \quad T_4=M_2,
\end{equation}
and, finally,
$$
A_1={N_1-N_2+M_2\over 2},\qquad A_2={N_2-N_3+M_1\over 2},
$$
$$ A_3={-N_3-N_4+M_2\over 2},\qquad A_4= {N_1+N_4+M_1\over 2}.
$$
We note that
\be\label{baldic}
\sum_{a=1}^4(R_a+U_a)=-4\textup{i}\quad{\rm and}\quad \sum_{a=1}^4(S_a+T_a)=0
\ee
and
\be\label{aaa}
A_1+A_2=A_3+A_4.
\ee

Racah coefficients \eqref{6jds} depend on 6 complex parameters (pairs of continuous and discrete
variables), whereas function \eqref{idsgamma} formally depends on 8 such parameters $(R_j,S_j)$, $(U_j,T_j)$,
$j=1,\ldots,4$, and the balancing condition leaves 7 independent quantities. However,
the structure of integration and summation in \eqref{idsgamma} allows shifting the integration and summation
variables by arbitrary constants (with an appropriate shift of the integration contour). Therefore, one of the
pairs of variables can be set equal to zero, which was done in \eqref{RU} and \eqref{T} by the choice $U_3=0$ and $T_3=0$.

\section{Relation to $6j$-symbols for the Faddeev modular double}

We consider Faddeev's quantum modular dilogarithm  $\gamma^{(2)}(y;\omega_1,\omega_2)$
\cite{Fad94,Fad95} also called the hyperbolic gamma function \cite{ruij},
\be
\gamma^{(2)}(u;\mathbf{\omega})= \gamma^{(2)}(u;\omega_1,\omega_2):=e^{-\frac{\pi\textup{i}}{2}
B_{2,2}(u;\mathbf{\omega}) } \gamma(u;\mathbf{\omega}),
\label{HGF}\ee
where $B_{2,2}$ is the second-order multiple Bernoulli polynomial
$$
B_{2,2}(u;\mathbf{\omega})=\frac{1}{\omega_1\omega_2}
\left((u-\frac{\omega_1+\omega_2}{2})^2-\frac{\omega_1^2+\omega_2^2}{12}\right)
$$
and
\be
\gamma(u;\mathbf{\omega}):= \frac{(\tilde q e^{2\pi \textup{i} \frac{u}{\omega_1}};\tilde q)_\infty}
{(e^{2\pi \textup{i} \frac{u}{\omega_2}};q)_\infty}
=\exp\left(-\int_{\mathbb{R}+\textup{i}0}\frac{e^{ux}}
{(1-e^{\omega_1 x})(1-e^{\omega_2 x})}\frac{dx}{x}\right).
\label{int_rep}\ee
This function obeys the first order difference equations
\be\label{hp1}
{\gamma^{(2)}(y+\omega_1;\omega_1,\omega_2)\over \gamma^{(2)}(y;\omega_1,\omega_2)}=2\sin{\pi y\over \omega_2},\quad
{\gamma^{(2)}(y+\omega_2;\omega_1,\omega_2)\over \gamma^{(2)}(y;\omega_1,\omega_2)}=2\sin{\pi y\over \omega_1}
\ee
and has the asymptotics \cite{Kharchev:2001rs}
\be\label{gamasym}
{\bf I:} \quad \stackreb{\lim}{y\to \infty}e^{{\pi\textup{i}\over 2}B_{2,2}(y,\omega_1,\omega_2)}\gamma^{(2)}(y;\omega_1,\omega_2)=1,
\quad \; {\rm arg}\;\omega_1<{\rm arg}\; y<{\rm arg}\;\omega_2+\pi,
\ee
\be\label{gamasym2}
{\bf II:} \quad \stackreb{\lim}{y\to \infty}e^{-{\pi\textup{i}\over 2}B_{2,2}(y,\omega_1,\omega_2)}\gamma^{(2)}(y;\omega_1,\omega_2)=1,
\quad \; {\rm arg}\;\omega_1-\pi<{\rm arg}\; y<{\rm arg}\;\omega_2.
\ee

In what follows, we shall a special degeneration limit
\begin{equation}\label{om1om22}
b:=\sqrt{\omega_1\over \omega_2}=\textup{i}+\delta, \quad \delta\to 0^+,
\end{equation}
in which case
\be\label{qlim}
Q=\omega_1+\omega_2=2\delta\sqrt{\omega_1\omega_2}+O(\delta^2).
\ee
In \cite{Sarkissian:2020ipg}, it was rigorously shown that in the limit (\ref{om1om22}), the estimate
\begin{equation}\label{gam2lim2}
\gamma^{(2)}(\textup{i}\sqrt{\omega_1\omega_2}(n+x\delta);\omega_1,\omega_2)\stackreb{=}{\delta\to 0^+} e^{\frac{\pi \textup{i}}{2}n^2} (4\pi\delta)^{\textup{i}x-1}{\bf \Gamma}(x,n),
\quad \sqrt{\omega_1\over \omega_2}=\textup{i}+\delta,
\end{equation}
holds uniformly on compacta,
where $n\in {\mathbb Z}, \, x\in {\mathbb C}$, and complex gamma function is defined in \eqref{Cgamma}.
Such a limit was qualitatively considered first in \cite{BMS}.

The $6j$-symbols corresponding to the principal unitary series representations of the Faddeev modular double
$sl_q(2,\mathbb{R})\times sl_{\tilde q}(2,\mathbb{R})$ \cite{fad:mod} were constructed
by Ponsot and Teschner in \cite{Ponsot:2000mt} ni the form of an explicit expression
\be\label{ponsot}
\big\{\,{}^{\alpha_1}_{\alpha_3}\,{}^{\alpha_2}_{{\alpha}_4}\,|\,{}^{\alpha_s}_{\alpha_t}\big\}_b=
{S_b(\alpha_s+\alpha_2-\alpha_1)S_b(\alpha_1+\alpha_t-\alpha_4)\over
S_b(\alpha_t+\alpha_2-\alpha_3)S_b(\alpha_3+\alpha_s-\alpha_4)}
|S_b(2\alpha_t)|^2J_b(\underline{\mu},\underline{\nu}),
\ee
where $S_b(z)=\gamma^{(2)}(z; b,b^{-1})$,
\be\label{Jb}
J_b(\underline{\mu},\underline{\nu})=\int_{-\textup{i}\infty}^{\textup{i}\infty}\prod_{a=1}^4S_b(\mu_a- z)S_b(\nu_a+ z)dz,
\ee
and
\begin{equation}\label{mu4}
\begin{aligned}
&\nu_1=\alpha_s+\alpha_1-\alpha_2,\\
&\nu_2=Q+\alpha_s-\alpha_1-\alpha_2,\\
&\nu_3=\alpha_s+\alpha_3-\alpha_4,\\
&\nu_4=Q+\alpha_s-\alpha_3-\alpha_4,
\end{aligned}\qquad
\begin{aligned}
&\mu_1=-Q-\alpha_s+\alpha_t+\alpha_4+\alpha_2,\\
&\mu_2=-\alpha_s-\alpha_t+\alpha_4+\alpha_2,\\
&\mu_3=Q-2\alpha_s,\\
&\mu_4=0,
\end{aligned}
\end{equation}
with $Q=b+b^{-1}$. We note that the parameters $\mu_a$ and $\nu_a$ satisfy the balancing condition
\be\label{bal}
\sum_{a=1}^4(\nu_a+\mu_a)=2Q.
\ee
Formally, function \eqref{Jb} depends on 7 independent complex variables, but one of the
parameters can be set equal to zero by shifting the integration variable, which was done in
 \eqref{mu4} by the choice $\mu_4=0$.

The factor $|S_b(2\alpha_t)|^2$ corresponds to the measure in the orthogonality relation
for $6j$-symbols. In principle, we can remove it from the definition \eqref{ponsot}
and lift the unitary principal series restrictions imposed on the parameters of this function
 $\alpha_{1,2,3,4}, \alpha_s,\alpha_t\in Q/2+\textup{i}\R$
(see the definition of the hyperbolic hypergeometric function $J_h(\underline{\mu},\underline{\nu})$
in \eqref{jhh} below, where we do not assume such a constraint).

Now we intend to show that formula (\ref{idsgamma}) can be derived from expression (\ref{ponsot})
in the limit $b\to \textup{i}$. First, we rewrite the last function in a slightly
different notation.
We shift the integration variable  $z\to z +Q-2\alpha_s$ and afterwards define new parameters
\begin{equation}
\begin{aligned}
&\alpha_s=\alpha_s'+Q/2,\\
&\alpha_t=-\alpha_t'+Q/2,
\end{aligned}\qquad
\begin{aligned}
&\alpha_1=-\alpha_1'+Q/2,\\
&\alpha_2=-\alpha_2'+Q/2,
\end{aligned}\qquad
\begin{aligned}
&\alpha_3=-\alpha_3'+Q/2,\\
&\alpha_4=\alpha_4'+Q/2.
\end{aligned}
\end{equation}
After the shifts, parameters $\nu_a$ and $\mu_a$ take the following form in terms of the  $\alpha'$-variables
\begin{equation}
\begin{aligned}
&\nu_1=Q/2-\alpha_s'-\alpha_1'+\alpha_2',\\
&\nu_2=Q/2-\alpha_s'+\alpha_1'+\alpha_2',\\
&\nu_3=Q/2-\alpha_s'-\alpha_3'-\alpha_4',\\
&\nu_4=Q/2-\alpha_s'+\alpha_3'-\alpha_4',
\end{aligned}\qquad
\begin{aligned}
&\mu_1=\alpha_s'-\alpha_t'+\alpha_4'-\alpha_2',\\
&\mu_2=\alpha_s'+\alpha_t'+\alpha_4'-\alpha_2',\\
&\mu_3=0,\\
&\mu_4=2\alpha_s'.
\end{aligned}
\label{munu}\end{equation}

For $\omega_1=b$, $\omega_2=b^{-1}$, relation (\ref{gam2lim2}) takes the form:
\be\label{limens}
S_{\textup{i}+\delta}(\textup{i}(n+x\delta))\stackreb{=}{\delta\to 0}
e^{\pi \textup{i} n^2\over 2}
(4\pi\delta)^{\textup{i}x-1}{\bf \Gamma}(x,n).
\ee

We parametrize our new $\alpha'$ variables and the integration variable in accordance with the form
of the argument of the $S_b$-function in formula (\ref{limens}):
\begin{equation}
\begin{aligned}
&\alpha_1'=\textup{i}(N_1/2+\sigma_1 \delta),\\
&\alpha_2'=\textup{i}(N_2/2+\sigma_2 \delta),
\end{aligned}\qquad
\begin{aligned}
&\alpha_3'=\textup{i}(N_3/2+\sigma_3 \delta),\\
&\alpha_4'=\textup{i}(N_4/2+\sigma_4 \delta),
\end{aligned}\qquad
\begin{aligned}
&\alpha_t'=\textup{i}(M_1/2+\rho_1 \delta),\\
&\alpha_s'=\textup{i}(M_2/2+\rho_2 \delta),
\end{aligned}\qquad
\begin{aligned}
&z=\textup{i}(-N-u \delta).
\end{aligned}
\end{equation}
In this parametrization, $N_k$ and $M_k$ are integers chosen such that their
linear combinations appearing in the expressions for $\mu_j$ and $\nu_j$ \eqref{munu} take
even values in order to be divisible by 2.

Now we can apply asymptotic formula \eqref{limens} and find that in the limit $\delta\to 0$,
the $6j$-symbols for Faddeev's modular double are converted to the complex $6j$-symbols for
the $SL(2,\C)$ group,
\bea
\big\{\,{}^{\alpha_1}_{\alpha_3}\,{}^{\alpha_2}_{{\alpha}_4}\,|\,{}^{\alpha_s}_{\alpha_t}\big\}_{\textup{i}+\delta}
\stackreb{=}{\delta\to 0} e^{{\pi\textup{i}\over 2}F}
{M_1^2+4\rho_1^2\over 16\pi^4\textup{i}\delta}\big\{\,{}^{\sigma_1,N_1}_{\sigma_3,N_3}\,{}^{\sigma_2,N_2}_{{\sigma}_4,N_4}\,|\,{}^{\rho_1,M_1}_{\rho_2,M_2}\big\},
\label{PTrel}\eea
where
\be
F=A_1^2-A_2^2-A_3^2+A_4^2+\sum_{a=1}^4 (S_a^2+T_a^2)+2(A_4-A_2).
\ee
Conditions (\ref{baldic}) and (\ref{aaa}) imply that $F$ is an even integer.
Therefore $e^{{\pi\textup{i}\over 2}F}$ is a plain sign factor.
The details of the computations can be found in Appendix A.
The proportionality coefficient in the right-hand side of relation \eqref{PTrel} without the diverging factor
$e^{{\pi\textup{i}\over 2}F}/\textup{i}\delta$ coincides with the $\rho$-function
defined in equation (28) of \cite{Derkachev:2019bcl}, which describes the measure weight function.
It emerged from the multiplier $|S_b(2\alpha_t)|^2$ in definition \eqref{ponsot}.

\section{Difference equations}

An elliptic analogue of the Euler-Gauss hypergeometric function $V(t_1,\ldots,t_8;p,q)$ was introduced in \cite{spi:theta},
\be\label{vfu}
V(t_1,\ldots,t_8;p,q)=\frac{(p;p)_\infty(q;q)_\infty}{4\pi\textup{i}}
\int_{\mathbb{T}}{\prod_{a=1}^8\Gamma(t_az; p,q)\Gamma(t_az^{-1}; p,q)
\over \Gamma(z^2; p,q)\Gamma(z^{-2}; p,q)}{dz\over z},
\ee
with the parameters $t_a$ satisfying the balancing condition
\be
 \prod_{a=1}^8t_a=p^2q^2.
\ee
Here, $\Gamma(z;p,q)$ is the elliptic gamma function defined as a double infinite product
\be
\Gamma(z;p,q)=\prod_{j,k=0}^{\infty}{1-z^{-1}p^{j+1}q^{k+1}\over 1-zp^jq^k},
\quad |p|,|q|<1,\quad z\in \mathbb{C}^*.
\ee
It satisfies the equations
\be
\Gamma(z;p,q)=\Gamma(z;q,p),
\ee
\be
\Gamma(qz;p,q)=\theta(z;p)\Gamma(z;q,p), \qquad
\Gamma(pz;p,q)=\theta(z;q)\Gamma(z;q,p),
\ee
where $\theta(z;p)$ is a short Jacobi theta-function
\be
\theta(z;q)=(z;q)_\infty (qz^{-1};q)_\infty, \quad (z;q)_\infty:=\prod_{j=0}^\infty (1-zq^j),
\ee
related to Jacobi's $\theta_1$-function as follows
\begin{eqnarray} \nonumber &&
 \theta_1(u|\tau)=-\theta_{11}(u)=-\sum_{\ell\in \mathbb Z+1/2}e^{\pi \textup{i} \tau \ell^2}
e^{2\pi \textup{i} \ell (u+1/2)}
\\  && \makebox[4em]{}
=\textup{i}q^{1/8}e^{-\pi {i} u}(q;q)_\infty\theta(e^{2\pi \textup{i}u};q).
\label{theta1}\end{eqnarray}

As shown in \cite{spi:thesis,spi:tmf}, the $V$-function satisfies a finite-difference equation, called the elliptic hypergeometric equation,
\be\label{elldif}
{\mathcal L}(\underline{t})(U(qt_6,q^{-1}t_7)-U(\underline{t}))+(t_6\leftrightarrow t_7)+U(\underline{t})=0,
\ee
where
\be
U(\underline{t})={V(t_1,\ldots,t_8;p,q)\over \Gamma(t_6 t_8^{\pm 1}; p,q)\Gamma(t_7 t_8^{\pm 1}; p,q)},
\ee
the first $U$-function containing parameters $qt_6, q^{-1}t_7$ instead of $t_6, t_7$, and
\bea\label{ba}
{\mathcal L}(\underline{t})={\theta\left({t_6\over qt_8};p\right)\theta\left(t_6 t_8; p\right)\theta\left({t_8\over t_6};p\right)
\over \theta\left({t_6\over t_7}; p\right)\theta\left({t_7\over qt_6};p\right)\theta\left({t_7t_6\over q};p\right)}
\prod_{k=1}^5{\theta\left({t_7t_k\over q};p\right)\over \theta(t_8t_k;p)}.
\eea

We recall the asymptotic relation \cite{ruij}
\be\label{rlim}
\Gamma(e^{-2\pi vy};e^{-2\pi v\omega_1},e^{-2\pi v\omega_2})
\stackreb{=}{ v\to 0}
e^{-\pi(2y-\omega_1-\omega_2)/12v\omega_1\omega_2}\gamma^{(2)}(y;\omega_1,\omega_2),
\ee
which was shown in \cite{rai:limits} to be uniform on compacta, and the asymptotics
\be
\theta(e^{-2\pi  vy}; e^{-2\pi  v\omega_1})
\stackreb{=}{v\to 0} e^{-{\pi\over 6v\omega_1}}2\sin{\pi y\over \omega_1}.
\ee
With their help, we can derive difference equations for the corresponding hyperbolic hypergeometric functions.

A hyperbolic analogue of the $V$-function (\ref{vfu}) appearing from it in the limit \eqref{rlim} is
\be\label{bi}
I_h(\underline{u})=\int_{-\textup{i}\infty}^{\textup{i}\infty}{\prod_{a=1}^8\gamma^{(2)}(u_a\pm z;\omega_1,\omega_2)\over
\gamma^{(2)}(\pm 2z)}\frac{dz}{2\textup{i}\sqrt{\omega_1\omega_2}},
\ee
with $u_a$ satisfying the condition:
\be\label{mu8}
\sum_{a=1}^8 u_a=2(\omega_1+\omega_2).
\ee

It can be shown \cite{BRS} that in the limit (\ref{rlim}), elliptic hypergeometric equation (\ref{elldif})
is converted into a difference equation for $I_h(\underline{u})$,
\be\label{br}
{\mathcal A}(u;\omega_1,\omega_2)(Y(u_6+\omega_2,u_7-\omega_2)-Y(u))+(u_6\leftrightarrow u_7)+Y(u)=0,
\ee
where
\bea\label{ba'} && \nonumber
{\mathcal A}(u;\omega_1,\omega_2)={\sin{{\pi\over\omega_1}}(u_6-u_8-\omega_2)\sin{{\pi\over \omega_1}}(u_6+u_8)\sin{{\pi\over \omega_1}}(u_8-u_6)
\over \sin{{\pi\over \omega_1}}(u_6-u_7)\sin{{\pi\over \omega_1}}(u_7-u_6-\omega_2)\sin{{\pi\over \omega_1}}(u_7+u_6-\omega_2)}
\\ && \makebox[8em]{} \times
\prod_{k=1}^5{\sin{{\pi\over \omega_1}}(u_7+u_k-\omega_2)\over \sin{{\pi\over \omega_1}}(u_8+u_k)}
\eea
and
\be
Y(\underline{u})={I_h(\underline{u})\over \gamma^{(2)}(u_6\pm u_8, u_7\pm u_8; \omega_1,\omega_2)}.
\ee
The function $I_h(u)$ is symmetric in the quasiperiods $\omega_1$ and $\omega_2$.
Therefore, we have a second difference equation
\be\label{br2}
{\mathcal A}(u;\omega_2,\omega_1)(Y(u_6+\omega_1,u_7-\omega_1)-Y(u))+(u_6\leftrightarrow u_7)+Y(u)=0.
\ee

A detailed consideration of the limiting transitions $\omega_1\to 0$ (for a fixed $\omega_2$) and
$\omega_1\to \pm \omega_2$ for the function $Y(u)$ and for the corresponding difference equations
was given in \cite{GSVS}. Here, we investigate a special degenerate case of this function:
our key object is now the integral
\be\label{ehh}
E_h(\underline{u})=\int_{-\textup{i}\infty}^{\textup{i}\infty}{\prod_{a=1}^6\gamma^{(2)}(u_a\pm z;\omega_1,\omega_2)\over
\gamma^{(2)}(\pm 2z;\omega_1,\omega_2)}\frac{dz}{2\textup{i}\sqrt{\omega_1\omega_2}}
\ee
without any balancing condition. For Re$(u_a)>0$, we can fix the integration contour as the imaginary axis.
For other values of parameters, the integration contour is of the Mellin-Barnes type, i.e.,
it should separate sequences of poles going to infinity to the left and right of the imaginary axis.

We now derive the difference equation for this function following from hyperbolic hypergeomertic
equation \eqref{br}. For this, we set $u_1=2(\omega_1+\omega_2)-\sum_{k=2}^8u_k$ and take the limit
$u_8\to \infty$ inside cone I \eqref{gamasym} (such that $u_1\to \infty$ inside cone II \eqref{gamasym2}).
Then using the corresponding asymptotics and renumbering the remaining parameters
$u_k\to u_{k-1}$, we obtain
\be\label{difeh}
{\mathcal B}(u;\omega_1,\omega_2)(E_h(u_5+\omega_2,u_6-\omega_2)-E_h(u))+(u_5\leftrightarrow u_6)+E_h(u)=0,
\ee
$$
{\mathcal B}(u;\omega_1,\omega_2)={\prod_{k=1}^4 \sin{{\pi\over \omega_1}}(u_6+u_k-\omega_2)
\over \sin{{\pi\over \omega_1}}(u_5-u_6)\sin{{\pi\over \omega_1}}(u_6-u_5-\omega_2)\sin{{\pi\over \omega_1}}(u_6+u_5-\omega_2)\sin{{\pi\over \omega_1}}(2Q-\sum_{i=1}^6u_i)}.
$$
This equation was also derived in \cite{BRS}. Despite of the absence of the
balancing condition, we can convert equation \eqref{difeh} into a finite-difference equation
of the second order in the variable $x$ by setting $u_5=c+x$ and $u_6=c-x$ for an arbitrary constant $c$.
Evidently, we also have the second equation
\be\label{difeh2}
{\mathcal B}(u;\omega_2,\omega_1)(E_h(u_5+\omega_1,u_6-\omega_1)-E_h(u))+(u_5\leftrightarrow u_6)+E_h(u)=0.
\ee

We can consider a different degeneration of equation (\ref{br}).
We reparametrize the variables $u_a$ in identity (\ref{br}) in the following asymmetric way:
\bea
&& u_a= \nu_a+\textup{i}\xi, \quad u_{a+4}= \mu_{a}-\textup{i}\xi,  \quad a=1,2,3,4.
\eea
Then the balancing condition takes the form
\be\label{ball}
\sum_{a=1}^4(\nu_a+\mu_a)=2(\omega_1+\omega_2).
\ee
In definition (\ref{bi}), we now shift the integration variable $z\to z-\textup{i}\xi$ and take the limit
$\xi\to -\infty$. From the above equations, we then obtain
\be\label{secdif}
{\mathcal D}(\underline{\mu},\underline{\nu};\omega_1,\omega_2)(U(\mu_2+\omega_2,\mu_3-\omega_2)
-U(\underline{\mu},\underline{\nu}))+(\mu_2\leftrightarrow \mu_3)+U(\underline{\mu},\underline{\nu})=0
\ee
and
\be\label{secdif2}
{\mathcal D}(\underline{\mu},\underline{\nu};\omega_2,\omega_1)(U(\mu_2+\omega_1,\mu_3-\omega_1)
-U(\underline{\mu},\underline{\nu}))+(\mu_2\leftrightarrow \mu_3)+U(\underline{\mu},\underline{\nu})=0,
\ee
where
\bea
{\mathcal D}(\underline{\mu},\underline{\nu};\omega_1,\omega_2)={\sin{{\pi\over\omega_1}}(\mu_2-\mu_4-\omega_2)
\sin{{\pi\over \omega_1}}(\mu_4-\mu_2)
\over \sin{{\pi\over \omega_1}}(\mu_2-\mu_3)\sin{{\pi\over \omega_1}}(\mu_3-\mu_2-\omega_2)}
\prod_{k=1}^4{\sin{{\pi\over \omega_1}}(\mu_3+\nu_k-\omega_2)\over \sin{{\pi\over \omega_1}}(\mu_4+\nu_k)},
\eea
\be\label{umunu}
 U(\underline{\mu},\underline{\nu})={J_h(\underline{\mu},\underline{\nu})\over \gamma^{(2)}(\mu_2- \mu_4, \mu_3- \mu_4; \omega_1,\omega_2)},
\ee
\be\label{jhh}
J_h(\underline{\mu},\underline{\nu})=\int_{-\textup{i}\infty}^{\textup{i}\infty}\prod_{a=1}^4\gamma^{(2)}(\mu_a- z;\omega_1,\omega_2)
\gamma^{(2)}(\nu_a+ z;\omega_1,\omega_2)\frac{dz}{\textup{i}\sqrt{\omega_1\omega_2}}
\ee
with the parameters $\mu_a$ and $\nu_a$ satisfying the condition (\ref{ball}).

In fact, equality (\ref{secdif}) is another form of equation (\ref{difeh}) because, as
established in \cite{BRS}, the integrals (\ref{ehh}) and (\ref{jhh}) are connected by the relation
\begin{eqnarray} \nonumber &&
J_h(\underline{\mu},\underline{\nu})=\prod_{a=1}^3\gamma^{(2)}(\mu_a+\nu_4;\omega_1,\omega_2)
\gamma^{(2)}(\nu_a+\mu_4;\omega_1,\omega_2)
\\ &&  \makebox[4em]{} \times
E_h(\mu_1+\eta,\mu_2+\eta,\mu_3+\eta,\nu_1-\eta,\nu_2-\eta,\nu_3-\eta),
\label{jheh} \end{eqnarray}
where
\be
2\eta=Q-\nu_4-\sum_{a=1}^3\mu_a.
\ee

\section{The $b\to 0$ reduction of difference equations}

We now investigate various degenerations of hyperbolic integrals \eqref{ehh} and \eqref{jhh}
corresponding to the limits $b\to 0$ and $b\to \textup{i}$. For the general difference equations \eqref{br} and \eqref{br2}, such degenerations are considered in \cite{GSVS}.
We start from the standard $\omega_1\to 0$ limit applied to equations \eqref{secdif} and \eqref{secdif2}.
Recall the following uniform asymptotics computed in \cite{ruij} (see also \cite{GSVS}):
\be
\gamma^{(2)}(\omega_1 x;\omega_1,\omega_2)\to {\Gamma(x)\over \sqrt{2\pi}}\left(\omega_2\over 2\pi\omega_1\right)^{{1\over 2}-x}
\ee
We plan to scale all the parameters according to the rule
\be\label{scpar}
\mu_k=\omega_1\beta_k, \quad \nu_k=\omega_1\gamma_k,\quad k=1,2,3,4, \quad {\rm and} \quad z=\omega_1 u.
\ee
However, we must first eliminate the $\omega_2$ term in the right-hand side of the balancing
condition \eqref{ball}. For this, we shift the parameters $\nu_{1,2}\to \nu_{1,2}+\omega_2$
(such a shift can be made differently, say, $\nu_1\to \nu_1+\omega_2, \mu_1\to \mu_1+\omega_2$,
but we limit ourselves to one example). Then, using the reflection rule
\be\label{refprop}
\gamma^{(2)}(x;\omega_1,\omega_2)\gamma^{(2)}(\omega_1+\omega_2-x;\omega_1,\omega_2)=1,
\ee
we rewrite integral \eqref{jhh} in the form:
\be\label{jhh3}
J_h(\underline{\mu},\underline{\nu})=\int_{-\textup{i}\infty}^{\textup{i}\infty}
{\prod_{i=1}^4\gamma^{(2)}(\mu_i- z;\omega_1,\omega_2)
\prod_{i=3}^4\gamma^{(2)}(\nu_i+ z;\omega_1,\omega_2)\over \prod_{i=1}^2\gamma^{(2)}(\omega_1-\nu_i-z;\omega_1,\omega_2)}\frac{dz}{\textup{i}\sqrt{\omega_1\omega_2}}.
\ee
Now we apply the scaling \eqref{scpar}, which leads to the balancing condition
\be
\sum_{k=1}^4(\beta_k+\gamma_k)=2
\ee
and in the limit $\omega_1\to 0$ obtain the relation
\be
J_h(\underline{\mu},\underline{\nu})\to {1\over (2\pi)^4}{\omega_2^2\over \omega_1} J_r(\underline{\beta},\underline{\gamma}),
\ee
where
\be
J_r(\underline{\beta},\underline{\gamma})=\int_{-\textup{i}\infty}^{\textup{i}\infty}{\prod_{i=1}^4\Gamma(\beta_i- u)
\prod_{i=3}^4\Gamma(\gamma_i+ u)\over \prod_{i=1}^2\Gamma(1-\gamma_i-u)}du.
\ee
Now it is straightforward to show  that equation \eqref{secdif2} in this limit is converted to
\be
{\mathcal D}(\underline{\beta},\underline{\gamma})({\mathcal J}_r(\beta_2+1,\beta_3-1,\gamma)-{\mathcal J}_r(\underline{\beta},\underline{\gamma}))+(\beta_2\leftrightarrow \beta_3)+{\mathcal J}_r(\underline{\beta},\underline{\gamma})=0,
\ee
where
\be
{\mathcal D}(\underline{\beta},\underline{\gamma})={(\beta_2-\beta_4-1)(\beta_4-\beta_2)
\over (\beta_2-\beta_3)(\beta_3-\beta_2-1)}
\prod_{k=1}^4{(\beta_3+\gamma_k-1)\over (\beta_4+\gamma_k)}
\label{D}\ee
and
\be
{\mathcal J}_r(\underline{\beta},\underline{\gamma})={J_r(\underline{\beta},\underline{\gamma})\over \Gamma(\beta_2-\beta_4)\Gamma(\beta_3-\beta_4)}.
\ee

In turn, equation \eqref{secdif} yields the following integral identity
\be
{e^{\pi\textup{i}(\beta_1-\beta_2)}\sin\pi(\beta_4-\beta_2)
\over \sin\pi(\beta_2-\beta_3)}
{\prod_{k=1}^2\sin\pi(\beta_3+\gamma_k)\over \prod_{k=3}^4\sin\pi(\beta_4+\gamma_k)}(\tilde{{\mathcal J}_r}(\underline{\beta},\underline{\gamma})-{\mathcal J}_r(\underline{\beta},\underline{\gamma}))+(\beta_2\leftrightarrow \beta_3)+{\mathcal J}_r(\underline{\beta},\underline{\gamma})=0,
\ee
\be
\tilde{{\mathcal J}_r}(\underline{\beta},\underline{\gamma})={\Gamma(1-\beta_2+\beta_4)\over \Gamma(\beta_3-\beta_4)}\int_{-\textup{i}\infty}^{\textup{i}\infty}e^{\pi\textup{i}(\beta_4-u)}
{\prod_{i=1,3,4}\Gamma(\beta_i- u)
\prod_{i=3}^4\Gamma(\gamma_i+ u)\over \Gamma(1-\beta_2+ u)\prod_{i=1}^2\Gamma(1-\gamma_i-u)}du.
\ee
The integral $\tilde{{\mathcal J}_r}$ converges in a rather cute way: the integrand falls off exponentially
fast $\propto u^{-2}e^{-2\pi\textup{i} u}$ for $u\to -\textup{i}\infty$, but
it has only a power suppression $\propto u^{-2}$  for $u\to +\textup{i}\infty$.

We can similarly simplify equation \eqref{difeh2} for $\omega_1\to 0$.
Here, we have no balancing condition, and after scaling the parameters $u_k$ and $z$ as
$u_k=\omega_1 \alpha_k,\, z=\omega_1 u$, integral \eqref{ehh} in the limit $\omega_1\to 0$
takes the form
\begin{eqnarray*} &&
E_h(\underline{u})\to \left({1\over 2\pi}\sqrt{{\omega_2\over\omega_1}}\right)^9\left({2\pi\omega_1\over \omega_2}\right)^{2\sum_{i=1}^6\alpha_i}E_r(\underline{\alpha}),
\\ \makebox[4em]{} &&
E_r(\underline{\alpha})=\int_{-\textup{i}\infty}^{\textup{i}\infty}{\prod_{i=1}^6\Gamma(\alpha_i\pm u)\over
\Gamma(\pm 2u)}{du\over 4\pi \textup{i}},
\end{eqnarray*}
and correspondingly  equation \eqref{difeh2} becomes
\be\label{difeh22}
{\mathcal C}(\underline{\alpha})(E_r(\alpha_5+1,\alpha_6-1)-E_r(\underline{\alpha}))
+(\alpha_5\leftrightarrow \alpha_6)+E_r(\underline{\alpha})=0,
\ee
where
\be
{\mathcal C}(\underline{\alpha})={\prod_{k=1}^4(\alpha_6+\alpha_k-1)
\over (\alpha_5-\alpha_6)(\alpha_6-\alpha_5-1)(\alpha_6+\alpha_5-1)(2-\sum_1^6\alpha_i)}.
\label{C}\ee

\section{The $b\to \textup{i}$ reduction}

In this section, we consider the $b=\textup{i}+\delta,\, \delta \to 0^+$ limit (\ref{om1om22}) using
asymptotic relation \eqref{gam2lim2}.
Namely, we apply this limit  to difference equation {\ref{secdif}) with the following parametrization
\be\label{zbi}
z=\textup{i}\sqrt{\omega_1\omega_2}(N+y\delta),
\quad \mu_a=\textup{i}\sqrt{\omega_1\omega_2}(n_a+s_a\delta),
\quad\nu_a=\textup{i}\sqrt{\omega_1\omega_2}(m_a+t_a\delta),
\ee
where $y,s_a, t_a\in\C$ and $N, n_a, m_a \in\Z+\ve,\, \ve=0, \frac{1}{2}$.
Evidently, one has the relations
\begin{eqnarray*} &&
\textup{i}\sqrt{\omega_1\omega_2}(N+y\delta)+\omega_2=\textup{i}\sqrt{\omega_1\omega_2}(N-1+(y-\textup{i})\delta)
+O(\delta^2),
\\ && \makebox[2em]{}
{\textup{i}\over \omega_1}\sqrt{\omega_1\omega_2}(N+y\delta)=N+\delta(\textup{i}N+y)+O(\delta^2).
\end{eqnarray*}
Balancing condition (\ref{ball}) together with relation (\ref{qlim}) imply that the parameters
$n_a$, $m_a$, $s_a$, and $t_a$ satisfy the constraints
\be
\sum_{a=1}^4( n_a+m_a)=0,\qquad \sum_{a=1}^4( s_a+t_a)=-4\textup{i}.
\ee
Now using the uniform estimate (\ref{gam2lim2}), we compute the asymptotics of function \eqref{umunu},
\begin{eqnarray}\label{unew} &&
U(\underline{\mu},\underline{\nu})\stackreb{\to}{\delta\to 0^+}
{(-1)^{2\ve(1+\sum_{a=1}^4m_a)}\over (4\pi\delta)^{1+\textup{i}(s_2+s_3-2s_4)}}
e^{{\pi\textup{i}\over 2}(n_1^2-n_4^2+\sum_{a=1}^4m_a^2)}{\mathcal
U}(\underline{s},\underline{n};\underline{t},\underline{m}),
\\ && \makebox[4em]{}
{\mathcal  U}(\underline{s},\underline{n};\underline{t},\underline{m})
={\mathcal{J}_{cr}(\underline{s},\underline{n};\underline{t},\underline{m})
\over {\bf \Gamma}(s_2-s_4,n_2-n_4){\bf \Gamma}(s_3-s_4,n_3-n_4)},
\end{eqnarray}
where the subscript ``cr'' means ``complex rational'' and
\be\label{jdiscret}
\mathcal{J}_{cr}(\underline{s},\underline{n};\underline{t},\underline{m})
=\frac{1}{4\pi}\sum_{N\in\Z+\ve}\int_{-\infty}^{\infty}\prod_{a=1}^4{\bf \Gamma}(s_a-y,n_a-N)
{\bf \Gamma}(t_a+y,m_a+N)dy.
\ee
It turns out that the choice $\ve=1/2$ reduces to the case $\ve=0$ after the replacement of $n_a\to n_a+\ve$,
or, equivalently, $m_a\to m_a-\ve$. Therefore, we can drop $\ve$ from the definition of $\mathcal{J}_{cr}$.
Evidently, function \eqref{jdiscret} coincides with the key element defining the $6j$-symbols
for the $SL(2,\mathbb{C})$ group \eqref{idsgamma}.

We note that the diverging prefactor in \eqref{unew} contains only the sum of $s_2$ and $s_3$, and does not contain $n_2$ and $n_3$. This implies  the following limit form of equation (\ref{secdif}):
\be\label{difjmn}
{\mathcal D}(\underline{\beta},\underline{\gamma})
({\mathcal U}(s_2-\textup{i},n_2-1,s_3+\textup{i},n_3+1)-{\mathcal U}(\underline{s},\underline{n},\underline{t},\underline{m}))+(s_2,n_2\leftrightarrow s_3,n_3)+{\mathcal U}(\underline{s},\underline{n},\underline{t},\underline{m})=0,
\ee
where the function ${\mathcal D}(\underline{\beta},\underline{\gamma})$ is fixed in \eqref{D}
with the parametrization
\be\label{bgk}
\beta_k={1\over 2}(\textup{i}s_k-n_k),\qquad \gamma_k={1\over 2}(\textup{i}t_k-m_k).
\ee
In Appendix B, we consider a simple particular case of equation \eqref{difjmn} when the function ${\mathcal U}$ is
reduced to a computable beta integral, which explicitly demonstrates essential properties of this equation.

We can compute the $\delta\to 0^+$ form of equation \eqref{secdif2} similarly. Using the relations
\begin{eqnarray*} &&
\textup{i}\sqrt{\omega_1\omega_2}(N+y\delta)+\omega_1=\textup{i}\sqrt{\omega_1\omega_2}(N+1+(y-\textup{i})\delta)
+O(\delta^2),
\\ && \makebox[2em]{}
{\textup{i}\over \omega_2}\sqrt{\omega_1\omega_2}(N+y\delta)=-N-\delta(y-\textup{i}N)+O(\delta^2),
\end{eqnarray*}
it can be shown that in the limit $\delta\to 0$, difference equation (\ref{secdif2}) becomes
\be\label{difjmn2}
{\mathcal D}(\underline{\beta},\underline{\gamma})({\mathcal U}(s_2-\textup{i},n_2+1,s_3+\textup{i},n_3-1)-{\mathcal U}(\underline{s},\underline{n},\underline{t},\underline{m}))+(s_2,n_2\leftrightarrow s_3,N_3)+{\mathcal U}(\underline{s},\underline{n},\underline{t},\underline{m})=0,
\ee
where we have now a different parametrization
\be\label{bgk2}
\beta_k={1\over 2}(\textup{i}s_k+n_k), \qquad \gamma_k={1\over 2}(\textup{i}t_k+m_k).
\ee

In a similar way, we can compute the $b\to \textup{i}$ limit forms of equations \eqref{difeh}
and \eqref{difeh2}. We parametrize $u_a$ and $z$ as
$$
z=\textup{i}\sqrt{\omega_1\omega_2}(N+y\delta),
\qquad u_a=\textup{i}\sqrt{\omega_1\omega_2}(l_a+p_a\delta),
$$
where $y,p_a\in\C$ and $N, l_a \in\Z+\ve,\, \ve=0, \frac{1}{2}$.
Then, in the limit $\delta\to 0$ integral \eqref{ehh} has the asymptotics
\be\label{ehlim}
E_h(\underline{u})\stackreb{\to}{\delta\to 0^+} (-1)^{2\ve+(2\ve-1)\sum_{k=1}^6l_k}(4\pi\delta)^{2\textup{i}\sum_{k=1}^6p_k-9} {\mathcal E}_{cr}(\underline{p},\underline{l}),
\ee
where
\be
{\mathcal E}_{cr}(\underline{p},\underline{l})={1\over 8\pi}
\sum_{N\in\Z+\ve}\int_{-\infty}^{\infty}(y^2+N^2)
\prod_{k=1}^6{\bf \Gamma}(p_k\pm y,l_k\pm N)dy.
\ee
We note that in contrust to function \eqref{jdiscret}, the variable $\ve$ cannot be removed from the
definition of ${\mathcal E}_{cr}$, and it plays a substantial role. The complex hypergeometric analogue
of relation \eqref{jheh} connecting ${\mathcal E}_{cr}$ and ${\mathcal J}_{cr}$ functions
is considered in the next section.

It is now straightforward to see that in the limit $b\to \textup{i}$,
equation  \eqref{difeh} reduces to the equality
\be\label{difeh22'}
{\mathcal C}(\underline{\alpha})({\mathcal E}_{cr}(p_5-\textup{i},l_5-1,p_6+\textup{i},l_6+1)-{\mathcal E}_{cr}(\underline{p},\underline{l}))+(p_5,l_5\leftrightarrow p_6,l_6)+{\mathcal E}_{cr}(\underline{p},\underline{l})=0,
\ee
where the potential ${\mathcal C}(\underline{\alpha})$ is defined in \eqref{C} with the parametrization
$\alpha_k={1\over 2}(-l_k+\textup{i}p_k)$. Equation \eqref{difeh2} takes the form
\be\label{difeh22''}
{\mathcal C}(\underline{\alpha})({\mathcal E}_{cr}(p_5-\textup{i},l_5+1,p_6+\textup{i},l_6-1)-{\mathcal E}_{cr}(\underline{p},\underline{l}))+(p_5,l_5\leftrightarrow p_6,l_6)+{\mathcal E}_{cr}(\underline{p},\underline{l})=0,
\ee
with $\alpha_k={1\over 2}(l_k+\textup{i}p_k)$.

\section{Additional symmetry transformation for complex $6j$-symbols}

The key symmetry transformation for the elliptic hypergeometric function $V(\underline{t})$ established
in \cite{spi:theta} generates a rich family of nontrivial relations between the $V$-functions with
different parametrizations of its arguments, as well as of its different limiting forms.
The generic hyperbolic degeneration preserves the structure of the initial relations.
However, if we takes some of the hyperbolic level parameters  to infinity,
some of the permutational symmetries are lost, and the number of different symmetry transformations
increases substantially. One such relations was described earlier in \eqref{jheh}.
By taking parameters to infinity in a slightly different way, another symmetry transformation for function $J_h(\underline{\mu},\underline{\nu})$ was derived in \cite{BRS},
\bea\label{ide1b}
&&
J_h(\underline{\mu},\underline{\nu})=\prod_{ j, k =1}^2\gamma^{(2)}(\mu_j+\nu_k;\omega_1,\omega_2)\prod_{ j, k =3}^4\gamma^{(2)}(\mu_j+\nu_k;\omega_1,\omega_2)
\\ \nonumber && \makebox[2em]{}
\times J_h(\mu_1+\eta,\mu_2+\eta,\mu_3-\eta,\mu_4-\eta,\nu_1+\eta,\nu_2+\eta,\nu_3-\eta,\nu_4-\eta),
\eea
where
\be
\eta={1\over 2}(\omega_1+\omega_2-\mu_1-\mu_2-\nu_1-\nu_2).
\ee

We now consider what kind of identity emerges from this relation in the limit $b\to \textup{i}$.
Using parametrization \eqref{zbi} and taking the limit $\delta\to 0^+$, we obtain the following
symmetry transformation:
\bea\label{ide1i}
&&
\mathcal{J}_{cr}(\underline{s},\underline{n};\underline{t},\underline{m})=e^{\pi\textup{i}A}\prod_{ j, k =1}^2{\bf \Gamma}(s_j+t_k,n_j+m_k)\prod_{ j, k =3}^4{\bf \Gamma}(s_j+t_k,n_j+m_k)
\\ \nonumber
&& \makebox[2em]{} \times \mathcal{J}_{cr}(s_1+Y,n_1+K,s_2+Y,n_2+K,s_3-Y,n_3-K,s_4-Y,n_4-K;
\\ \nonumber &&  \makebox[4em]{}
t_1+Y,m_1+K,t_2+Y,m_2+K,t_3-Y,m_3-K,t_4-Y,m_4-K),
\\ \nonumber && \makebox[4em]{}
K=-{n_1+n_2+m_1+m_2\over 2},\quad\quad Y=-{s_1+s_2+t_1+t_2+2\textup{i}\over 2},
\\ &&  \makebox[-1em]{}
A=(n_1+n_2)(m_1+m_2)+(n_3+n_4)(m_3+m_4) +2(\ve+\lambda)\Big(1+\sum_{a=1}^4 m_a\Big).
 \nonumber\end{eqnarray}
In this relation, we have two discrete parameters $\ve, \lambda=0,\frac{1}{2}$ in the values of summation variables in $\mathcal{J}_{cr}$ in the respective left- and right-hand sides.
If $K$ is an integer, then  $\ve=\lambda$, and if $K$ is a half-integer, then $\ve\neq \lambda$.
However, as we have mentioned already, the variable $\ve$ can be removed from the $\mathcal{J}_{cr}$ function
in the left-hand side by the shifts $n_k\to n_k+\ve$, $m_k\to m_k-\ve$. Therefore we can assume that
$\ve=0$, and then in the right-hand side we have $\lambda=0$ for integer $K$ and $\lambda=\frac{1}{2}$
for half-integer $K$.

We now consider similar consequences for identity (\ref{jheh}). Using again parametrization \eqref{zbi}
and taking the limit $\delta\to 0^+$, we find a connection between the $\mathcal{J}_{cr}$ and $\mathcal{E}_{cr}$
functions
\be
\mathcal{J}_{cr}(\underline{s},\underline{n};\underline{t},\underline{m})=e^{\pi\textup{i} A}\prod_{a=1}^3{\bf \Gamma}(s_a+t_4,n_a+m_4){\bf \Gamma}(t_a+s_4,m_a+n_4)
\mathcal{E}_{cr}(\underline{s}+Z,\underline{n}+L,\underline{t}-Z,\underline{m}-L),
\label{JE}\ee
where
\begin{eqnarray} \nonumber &&
\mathcal{E}_{cr}(\underline{s}+Z,\underline{n}+L,\underline{t}-Z,\underline{m}-L)=\sum_{N\in\Z+\lambda }\int_{y\in\R} (y^2+N^2)
\\ && \makebox[4em]{} \times
\prod_{a=1}^3{\bf\Gamma}(s_a+Z\pm y,n_a+L\pm N){\bf\Gamma}(t_a-Z\pm y,m_a-L\pm N)dy,
\\ && \makebox[4em]{}
L=-{1\over 2}(m_4+\sum_{a=1}^3 n_a), \quad  Z=-{1\over 2}(t_4+2\textup{i}+\sum_{a=1}^3 s_a),
 \nonumber \\ && \makebox[4em]{}
A=2L^2-\Big(\sum_{a=1}^4n_a\Big)-2n_4m_4-\lambda  +2\ve \Big(1+\sum_{a=1}^4 m_a\Big).
\nonumber \end{eqnarray}
Since $\mathcal{E}_{cr}(\underline{s})$ evidently depends on 6 independent complex parameters,
the derived identity shows that this is true for the $\mathcal{J}_{cr}$ function as well.

In relation \eqref{JE}  we have two discrete parameters $\ve, \lambda=0,\frac{1}{2}$.
If $L$ is an integer, then $\ve=\lambda$, and if $L$ is a half-integer, then $\ve\neq \lambda$.
Again, without loss of generality, we can assume that $n_k, m_k\in\Z$, i.e. $\ve=0$, and  then
$\lambda=0$ for integer $L$, whereas $\lambda=\frac{1}{2}$ for half-integer $L$.
We conclude that both branches of the function $\mathcal{E}_{cr}$ are related to
$\mathcal{J}_{cr}$ and thus together they provide, in a rather intricate way,
a new unexpected integral representation of the $6j$-symbols for the $SL(2,\C)$ group,
a representation structurally different from the Mellin-Barnes representation derived
in \cite{Derkachev:2019bcl,Ismag2}.

\appendix

\section{Details of the $b\to\textup{i}$ asymptotics computations}

In this appendix, we give asymptotic estimates for hyperbolic
gamma functions in the limit $\delta\to 0^+$, which in combination reduce the $6j$-symbols
for the Faddeev modular double \eqref{ponsot} to $6j$-symbols for the $SL(2,\C)$ group \eqref{idsgamma}.
The constraints on parameters were indicated in the main body of the paper:
\be
S_{\textup{i}+\delta}(\mu_a- z)\to e^{{\pi\textup{i}\over 2}(T_a+N)^2}(4\pi\delta)^{\textup{i}(U_a+u )-1}{\bf \Gamma}(U_a+u ,T_a+N),
\ee
\be
S_{\textup{i}+\delta}(\nu_a+z)\to e^{{\pi\textup{i}\over 2}(S_a-N)^2}(4\pi\delta)^{\textup{i}(R_a-u )-1}{\bf \Gamma}(R_a-u ,S_a-N),
\ee
\bea
S_{\textup{i}+\delta}(\alpha_s+\alpha_2-\alpha_1)\to  e^{{\pi \textup{i}\over 2} A_1^2}
(4\pi\delta)^{\textup{i}(\sigma_1-\sigma_2+\rho_2)}{\bf \Gamma}(\sigma_1-\sigma_2+\rho_2-\textup{i},A_1),
\eea
\be
S_{\textup{i}+\delta}(\alpha_1+\alpha_t-\alpha_4)\to e^{{\pi \textup{i}\over 2} A_4^2}
{(4\pi\delta)^{\textup{i}(-\sigma_1-\sigma_4-\rho_1)}(-1)^{A_4}\over {\bf \Gamma}(\sigma_1+\sigma_4+\rho_1-\textup{i},A_4)},
\ee
\be
S_{\textup{i}+\delta}(\alpha_2+\alpha_t-\alpha_3)\to e^{{\pi \textup{i}\over 2} A_2^2}
{(4\pi\delta)^{\textup{i}(-\sigma_2+\sigma_3-\rho_1)}(-1)^{A_2}\over {\bf \Gamma}(\sigma_2-\sigma_3+\rho_1-\textup{i},A_2)},
\ee
\be
S_{\textup{i}+\delta}(\alpha_3+\alpha_s-\alpha_4)\to e^{{\pi \textup{i}\over 2} A_3^2}
(4\pi\delta)^{\textup{i}(-\sigma_3-\sigma_4+\rho_2)}{\bf \Gamma}(-\sigma_3-\sigma_4+\rho_2-\textup{i},A_3),
\ee
\be
|S_{\textup{i}+\delta}(2\alpha_t)|^2\to (4\pi\delta)^2{M_1^2+4\rho_1^2\over 4}.
\ee

\section{A check of equation (\ref{difjmn})}

We verify difference equation (\ref{difjmn}) for some particular choices of the
parameters. Namely, we remark that integral (\ref{jhh}) can be explicitly computed if, say,
$\mu_4+\nu_4=Q$ and $\sum_a^3(\mu_a+\nu_a)=Q$. In this case,
$\gamma^{(2)}(\mu_4- z;\omega_1,\omega_2)\gamma^{(2)}(\nu_4+ z;\omega_1,\omega_2)=1$ and
\be
J_h(\underline{\mu},\underline{\nu})
=\int_{-\textup{i}\infty}^{\textup{i}\infty}\prod_{a=1}^3\gamma^{(2)}(\mu_a- z;\omega_1,\omega_2)
\gamma^{(2)}(\nu_a+ z;\omega_1,\omega_2)\frac{dz}{\textup{i}\sqrt{\omega_1\omega_2}}
=\prod_{a,b=1}^3 \gamma^{(2)}(\mu_a+\nu_b;\omega_1,\omega_2).
\ee
Similarly, if $n_4+m_4=0$  and $s_4+t_4=-2\textup{i}$,
then ${\bf \Gamma}(s_4-y,n_4-N) {\bf \Gamma}(t_4+y,m_4+N)=(-1)^{N-n_4}$.
As a result, for
$$
\sum_{a=1}^3( n_a+m_a)=0,\, \sum_{a=1}^3( s_a+t_a)=-2\textup{i}
$$
we have \cite{Sarkissian:2020ipg}
\bea \nonumber &&
\mathcal{J}_{cr}(\underline{s},\underline{n};\underline{t},\underline{m})
=\frac{1}{4\pi}
\sum_{N\in\Z}(-1)^{N-n_4}\int_{-\infty}^{\infty}\prod_{a=1}^3{\bf \Gamma}(s_a-y,n_a-N)
{\bf \Gamma}(t_a+y,m_a+N)dy
\\ && \makebox[6em]{}
=(-1)^{\sum_{a=1}^4 n_a}F(\underline{s},\underline{n};\underline{t},\underline{m}),
\label{jdiscret'}\eea
where
\be\label{funcgam}
F(\underline{s},\underline{n};\underline{t},\underline{m})
=\prod_{a,b=1}^3 {\bf \Gamma}(s_a+t_b, n_a+m_b).
\ee
From the equations
\be\label{gnm}
{\bf \Gamma}(x-\textup{i},n-1)=\frac{{\bf \Gamma}(x,n)(n-\textup{i}x)}{2},\qquad
\quad {\bf \Gamma}(x+\textup{i},n+1)={{\bf \Gamma}(x,n)\over n/2-\textup{i}x/2+1}
\ee
and parametrization (\ref{bgk}) we find
\begin{eqnarray}\label{ftran} && \makebox[2em]{}
F(s_2-\textup{i},n_2-1,s_3+\textup{i},n_3+1)=\prod_{a=1}^3\frac{\beta_2+\gamma_a}
{\beta_3+\gamma_a-1}F(\underline{s},\underline{n};\underline{t},\underline{m}),
\\ \label{gamtran} &&
{{\bf \Gamma}(s_2-s_4,n_2-n_4)
{\bf \Gamma}(s_3-s_4,n_3-n_4)\over {\bf \Gamma}(s_2-\textup{i}-s_4,n_2-1-n_4)
{\bf \Gamma}(s_3+\textup{i}-s_4,n_3+1-n_4)}={\beta_3-\beta_4-1\over \beta_2-\beta_4}.
\end{eqnarray}

As a result, imposing the above constraints on the parameters and relations (\ref{ftran}),
(\ref{gamtran}), we obtain the algebraic equation
\bea  \nonumber &&
(\beta_2-\beta_4-1)(\beta_3-\beta_4)(\beta_3-\beta_2+1)\Big[
\prod_{k=1}^3(\beta_2+\gamma_k)(\beta_3-\beta_4-1)+ \prod_{k=1}^3(\beta_3+\gamma_k-1)(\beta_4-\beta_2)\Big]
\\  \nonumber &&
+(\beta_3-\beta_4-1)(\beta_2-\beta_4)(\beta_3-\beta_2-1)\Big[
\prod_{k=1}^3(\beta_3+\gamma_k)(\beta_2-\beta_4-1)+ \prod_{k=1}^3(\beta_2+\gamma_k-1)(\beta_4-\beta_3)\Big]
\\ && \makebox[4em]{}
-(\beta_2-\beta_3)(\beta_3-\beta_2+1)(\beta_3-\beta_2-1)\prod_{k=1}^3(\beta_4+\gamma_k)=0,
\label{eqdif} \eea
which is identically satisfied.
Taking the limit $\beta_4\to \infty$ in (\ref{eqdif}), we obtain
\bea
&& \makebox[2em]{}
(\beta_3-\beta_2+1)\Big(\prod_{k=1}^3(\beta_2+\gamma_k))-\prod_{k=1}^3(\beta_3+\gamma_k-1)\Big)
+(\beta_3-\beta_2-1)
\\ \nonumber && \times
\Big(\prod_{k=1}^3(\beta_3+\gamma_k)-\prod_{k=1}^3(\beta_2+\gamma_k-1)\Big)
+(\beta_3-\beta_2+1)(\beta_3-\beta_2-1)(\beta_2-\beta_3)=0,
\eea
which leads to a difference equation for function {\ref{funcgam}),
\be
{\mathcal V}(\underline{s},\underline{n};\underline{t},\underline{m})
(F(s_2-\textup{i},n_2-1,s_3+\textup{i},n_3+1)-F(s_2,n_2,s_3,n_3))+(2\leftrightarrow 3)+F(s_2,n_2,s_3,n_3)=0,
\ee
where
\bea
{\mathcal V}(\underline{s},\underline{n};\underline{t},\underline{m})= {\prod_{k=1}^3(\beta_3+\gamma_k-1)
\over (\beta_3-\beta_2-1)(\beta_2-\beta_3)}.
\eea

%\bigskip
%
%{\bf Conflict of Interest:} The authors declare that they have no conflicts of interest.

\end{document}